\documentclass{jaa}
\usepackage{graphicx}

\begin{document}\sloppy

%%paper title
%%For line breaks \\ can be used within title 
\title{On the cosmic origin of Fluorine}

%%author names are separated by comma (,) 
%%use \and before the last author name 
%%use a * along with the number separated by comma
%% for the  author for correspondence
%%\textsuperscript{number} is used for affiliation
%%\affilOne, \affilTwo etc., upto \affilTwentyfive is possible
%%Please note the first letter after \affil is capitalised in the command
%%

\author{Nils Ryde\textsuperscript{1,*}}
\affilOne{\textsuperscript{1}Lund Observatory, Department of Astronomy and Theoretical Physics, 
Lund University, Box 43, SE-221 00 Lund, Sweden. \\}

%%escape two column mode for title, affiliation and abstract
%%by giving \twocolumn command as shown

\twocolumn[{

\maketitle

%%include \corres to print the corresponding author Email id
\corres{ryde@astro.lu.se}

%%include \msinfo for
%%manuscript information such as
%%received, revised and accepted dates
%%
\msinfo{07 Sep. 2020}{20 Sep. 2020}

%%abstract
\begin{abstract}
The cosmic origin of fluorine, the ninth element of the Periodic Table, is still under debate. The reason for this fact is the large difficulties in observing stellar diagnostic lines, which can be used for the determination of the fluorine abundance in stars. Here we discuss some recent work on the chemical evolution of fluorine in the Milky Way and discuss the main contributors to the cosmic budget of fluorine. 
\end{abstract}

%%insert keywords separated by 3 hyphens using \keywords{words}
\keywords{abundances---fluorine---red giants.}

}]
%%close the twocolumn escape here

%%include \doinum{number}for the DOI number in the header
%%include \volnum{number} for the volume number in the header
%%include \year{yyyy} for  year of publication in the header
%%include \pgrange{num--num} page range of article in the header
%%include \artcitid{num} for the article citation id
%%include \lp to print last page of the article
%%include \setcounter{page}{pagenum} for the exact starting page of the article

\doinum{12.3456/s78910-011-012-3}
\artcitid{\#\#\#\#}
\volnum{000}
\year{0000}
\pgrange{1--}
\setcounter{page}{1}
\lp{1}

\section{Introduction}
Since 2019 was the 150$^\mathrm{th}$ anniversary of the Periodic Table of Chemical Elements, UNESCO proclaimed it as the 'International Year of the Periodic Table of Chemical Elements'. However, it was not until 2016, that the Periodic Table was completed up to period 7, with the identification of the element Oganesson with atomic number 118. 

Much astrophysical work has been devoted to understanding the cosmic origin of the elements, with large success for a range of elements. However, one element that has escaped a full understanding of its origin is fluorine, which has an atomic number of 9. The reason for this lack of understanding is the very few measurements done of the abundance evolution of fluorine in the Universe.

Fluorine is interesting since it is the most electronegative element, it is extremely chemically reactive, as well as  very nuclear reactive with p (H) and alpha (He) nuclei. On Earth fluorine can be found in many  minerals (such as the colourful Calcium fluorite, a.k.a fluorspar) and chemical compounds (such as the hazardous liquid HF, hydrogen fluoride). Fluorine was first isolated in 1888 by Henri Moissan, which earned him the Nobel Prize in Chemistry in 1906. 

The cosmic abundance of fluorine is several orders of magnitude lower than that of the neighbouring elements in the Periodic Table. It is only the 24th most common element in the Universe. It is thus clear that stellar nucleosynthesis processes must by-pass it. Indeed, the nuclear reaction cross sections F(p, and F(alpha, in these processes, are very high, which leads to the destruction of much of the newly formed fluorine. 

The main cosmic formation processes and sites for the production of fluorine is reviewed in detail in Ryde et al. (2020), and include: 

\begin{enumerate}
\item  in massive stars: The contribution from non-rotating stars, from conventional Type II supernovae explosions (SNIIe), is negligible. However, rapidly rotating massive stars can produce primary
fluorine from $^{14}$N, via proton and alpha captures (Prantzos
et al. 2018). Also, other processes have been suggested, such as the $\nu$-process in SNIIe. In this process neutrino-induced spallation reactions with $^{20}$Ne in the
expelled shell, can form fluorine. Also, contributions from  Wolf-Rayet stars during the core-helium burning phase, and
with the subsequent ejection through the strong, metal-line
driven wind, have been suggested, but is highly uncertain. 
\item  in low-mass, thermal-pulsing AGB stars ($2-4\,M_\odot$) fluorine is formed through a chain
of reactions involving neutrons and protons. Fluorine is expelled through stellar winds. These AGB
stars also form s-process elements like Ce. 
\item  in novae where fluorine is formed from reactions starting with proton captures on $^{17}$O nuclei. 
\end{enumerate}

So, which processes were the ones that succeeded in producing fluorine in the Universe, so that it survived to contribute to the comic build up of the element?  Here, we will discuss how to find this out and what has been done lately in this respect.

\begin{figure*}[!t]
\includegraphics{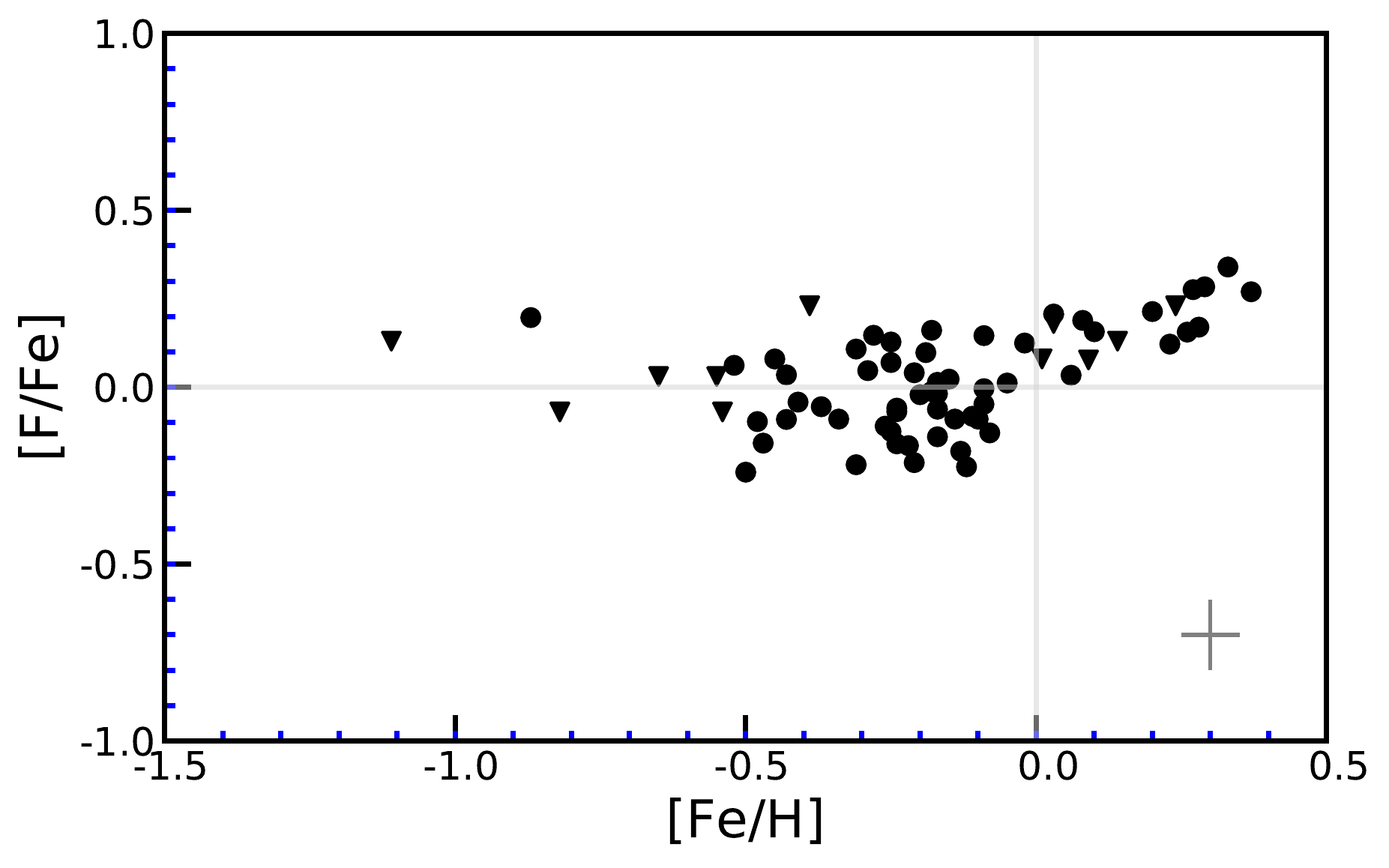}
\caption{[F/Fe] ratio as a function of metallicity is shown for the stars
observed in Ryde et al. (2020). Typical errorbars are indicated in the lower right corner.}\label{figOne}
\end{figure*}

\section{Determination of the cosmic origin of fluorine}

No useful atomic lines are available for use in an abundance determination in cool
stars. In hot stars it is possible to observe highly ionized far-UV lines (Werner et al. 2005)  and highly excited optical F{\sc i} lines (Pandey 2006; Pandey et al. 2008). The only readily useful diagnostics are instead the lines from the hydrogen-fluoride molecule, lines that lie in the $2.1–2.4$
(vibration-rotational lines) and $8–13\,\mu$m (pure rotational lines) regions. Only one single useful line, namely the HF(v = 1 - 0) R9 line at $\lambda_\mathrm{air}= 23\,358$ \AA, has been used in most investigations in the literature. We note that there has been a confusion about zero-point energy of the energy levels of the HF molecule in the literature, see J\"onsson et al. (2014). 

The way the fluorine abundance is retrieved is, in general,  by comparing synthetic spectra with observed spectra. In order to calculate these synthetic spectra, the stellar parameters of the star observed has to be known accurately. Especially, the effective temperatures of the stars have to be known accurately since the HF lines are very temperature sensitive. For example, in the recent study by Ryde et al. (2020) special care was taken to analyse a  homogeneous set of stars with well-determined stellar parameters. In that study, they determined the fluorine abundances for 51 stars spanning a range of metallicities, from [Fe/H]$
=-1.1$ to $+0.4$. The stars were observed with high-resolution, near-IR spectrographs at $R\sim 45000$. The accurate and precise stellar parameters were determined by the method described in J\"onsson et al. (2017). In this method the stellar parameters are determined simultaneously by fitting unsaturated and unblended lines  from Fe I, Fe II, and Ca I lines and  $\log$g sensitive Ca I wings. 

\section{The Ryde et al. (2020) study}

Of the 51 stars in the Ryde et al. (2020)  study, 41 stars were observed with the Phoenix spectrograph (Hinkle et al. 1998, 2003) mounted on the 4 m Mayall telescope at Kitt
Peak National Observatory (KPNO). Another 10 stars were observed with the IGRINS spectrograph at the 4.3 m Lowell Discovery Telescope (LDT; Mace et al. 2018), or on the 2.7 m Harlan J. Smith Telescope at McDonald Observatory (Mace et al.
2016). All in all, there were 51 spectra useful for the determination of the stellar fluorine abundances. 

From these spectra, Ryde et al. (2020) determined not only the fluorine abundance, but also those of oxygen and Ce. In Figure \ref{figOne} the [F/Fe] trend is shown as a function of the metallicity, [Fe/H].  The stellar abundances are
normalized to the solar value, which is very uncertain. Since there is
no detectable HF in the solar photosphere, the solar
abundance value is determined from meteoritic measurements
($A_\odot(F) = 4.43$; Lodders 2003).  

For metallicities below solar, a flat relation is found, with a smaller scatter compared to earlier results. At supersolar metallicities, a clear increasing relation is found. Furthermore, Ryde et al. (2020) find that the fluorine increase with metallicity shows a clear secondary behavior. 

They also find that the [F/Ce] ratio is relatively flat for $-0.6 < \mathrm{[Fe/H]} < 0$. For two s-process element-enhanced giants at [Fe/H]$ < -0.8$, they do not detect an elevated fluorine abundance. Based on their full data set, also their oxygen abundances which are compared to that of fluorine, Ryde et al. (2020) conclude that several major processes must be at play for the cosmic budget of fluorine over time. At lowest metallicities massive stars are most important and for $-0.6 < \mathrm{[Fe H]} < 0$ the asymptotic giant branch (AGB) star contribution must be large. At supersolar metallicities there seems be a need for processes with increasing yields with metallicity. The origins of the latter, and whether or not Wolf–Rayet stars and/or novae could contribute at supersolar metallicities, is
currently not known.

\section{The Grisoni et al. (2020) study}

In order to investigate the origin and evolution of fluorine based on the data from the Ryde et al. (2020) study, Grison et al. (2020) investigated the trends with detailed chemical evolution models. They indeed find that rotating massive stars should be  major contributors to the fluorine budget setting a plateau in the fluorine abundance trends below [Fe/H]$=-0.5$. For the increase at higher metallicities or later times, a contribution from low-mass stars is needed, in the form of single AGB stars and/or novae.

%%Use table* environment to get the table spanning both the columns

%\begin{table*}[htb]
%\tabularfont
%\caption{Caption text here}\label{secondTable}
%\begin{tabular}{lccccccccccccr}
%\topline
%\textbf{head1}&\multicolumn{11}{c}{\textbf{head2}}&\textbf{head3}\\
%\midline
%one& two &three&four&five&six&seven&eight&nine&ten&eleven&twelve&thirteen\\
%1&2&3&4&5&6&7&8&9&10&11&12&13\\
%aaa&bbbb&cccc&ddddd&eee&ffff&ggggg&hhhhhhhh&iiii&kkkkkk&hhh&jjjjjj&lllll\\
%\hline
%\end{tabular}
%\tablenotes{Table footnote here. Table spanning both the columns.}
%\end{table*}

%%An example of a figure

%%An example of a double column figure
%%Use figure* environment

%\begin{figure*}
%\centering\includegraphics[height=.15\textheight]{fig1.eps}
%\caption{caption spanning two columns}
%\centering\includegraphics[height=.25\textheight]{fig1.eps}
%\caption{caption here}
%\end{figure*}

\section{Conclusion}

The cosmic origin of the 9th element in the periodic table, fluorine, is still enigmatic. The light elements are the first to form in stellar nucleosynthetic processes, but fluorine is easily destroyed by nuclear reactions with the ubiquitous protons and alpha particles. Therefore, the cosmic abundance of fluorine is much lower that the neighbouring elements in the Periodic Table. A few observational studies have recently appeared. For example, the Ryde et al. (2020) study showed clearly a flat [F/Fe] trend at a solar value, with increasing fluorine abundance ratios at super solar metallicities. Their results also indicate that the fluorine slope shows a clear secondary behavior. Furthermore, they show that the [F/Ce] ratio is flat for $-0.6<\mathrm{[Fe/H]}<0.0$. Together with the fact that two metal-poor [Fe/H]$<-0.8$, Ce-enhanced stars do not have enhanced fluorine ratios, 
AGB stars cannot be the main contributor at these metallicities. Their conclusion is that several processes are needed for different metallicities: from massive, rotating stars at the lowest metallicities, to AGB stars at solar [Fe/H], to a processes with metallicity dependent yields at super solar metallcities. The Galactic Chemical Evolution models presented in Grisoni et al. (2020) agree with these conclusions and they show that there might be needed a contribution of novae for the production of fluorine in the Universe.

%%Use section* for acknowledgements
\section*{Acknowledgements}
N.R. acknowledges support from the Swedish Research Council, VR (project numbers 621-2014-5640), the Royal Physiographic Society in Lund through  the Stiftelse Walter Gyllenbergs fond and M{\"a}rta och Erik Holmbergs donation, the Crafoord Foundation, Stiftelsen Olle Engkvist Byggm\"astare, and Ruth och Nils-Erik Stenb\"acks stiftelse.
This work used the Immersion Grating Infrared spectrograph (IGRINS) that was developed under a collaboration between the University of Texas at Austin and the Korea Astronomy and Space Science Institute (KASI) with the financial support of the US National Science Foundation under grants AST-1229522 and AST-1702267, of the McDonald Observatory of the University of Texas at Austin, and of the Korean GMT Project of KASI.
\vspace{-1em}

%%use \balance somewhere in the left column of the last page to balance the two columns in the end page

%%References section


\begin{theunbibliography}{} 
\vspace{-1.5em}
\bibitem{latexcompanion} 
Grisoni, V., Romano, D., Spitoni, E., et al. 2020, MNRAS, ArXiv 2008.00812
\bibitem{latexcompanion} 
Hinkle, K. H., Blum, R. D., Joyce, R. R., et al. 2003, in Proc. SPIE, Vol. 4834, Discoveries and Research Prospects from 6- to 10-Meter-Class Telescopes II., ed. P. Guhathakurta, 353
\bibitem{latexcompanion} 
Hinkle, K. H., Cuberly, R. W., Gaughan, N. A., et al. 1998, SPIE, 3354, 810
\bibitem{latexcompanion}
J\"onsson, H., Ryde, N., Harper, G. M., Richter, M. J., Hinkle, K. H. 2014a, ApJL, 789, L41
\bibitem{latexcompanion}
J\"onsson, H., Ryde, N., Nordlander, T., et al. 2017a, A\&A,598, A100
\bibitem{latexcompanion} 
Lodders, K. 2003, ApJ, 591, 1220
\bibitem{latexcompanion} 
Mace, G., Sokal, K., Lee, J.-J., et al. 2018, in Society of Photo-Optical Instrumentation Engineers (SPIE) Conference Series, Vol. 10702, Proc. SPIE, 107020Q
\bibitem{latexcompanion} 
Mace, G., Kim, H., Jaffe, D. T., et al. 2016, Society of Photo-Optical Instrumentation Engineers (SPIE) Conference Series, Vol. 9908, 300 nights of science with IGRINS at McDonald Observatory, 99080C
\bibitem{latexcompanion} 
Pandey, G. 2006, ApJL, 648, L143
\bibitem{latexcompanion}
Pandey, G., Lambert, D. L., Kameswara Rao, N. 2008, ApJ, 674
\bibitem{latexcompanion} 
Prantzos, N., Abia, C., Limongi, M., Chieffi, A., Cristallo, S. 2018,
MNRAS, 476, 3432
\bibitem{latexcompanion} 
Ryde, N., J\"onsson, H., Mace, G. et al.  2020, ApJ, 893, 37 
\bibitem{latexcompanion} 
Werner, K., Rauch, T., Kruk, J. W. 2005, A\&A, 433, 641
\end{theunbibliography}
\end{document}